\documentclass[final,1p,times]{elsarticle}

\usepackage{amssymb}
\usepackage{amsmath}

\journal{Annals of Physics}

\begin{document}

\begin{frontmatter}



\title{Frame-dependent coherence of a quantum state}


\author{Nicolae Cotfas} 

\affiliation{organization={Faculty of Physics, University of Bucharest},
            addressline={P.O.Box MG-11}, 
            city={Bucharest},
            postcode={077125}, 
            country={Romania}}
 \ead{nicolae.cotfas@g.unibuc.ro}
\ead{ncotfas@yahoo.com}
 \ead[url]{https://unibuc.ro/user/nicolae.cotfas/}
\begin{abstract}
The tight frames can be regarded as a particular case of POVMs (positive operator-valued measures describing generalized measurements), namely the case when all the operators are rank-one.
Each orthonormal basis is a tight frame, and  every tight frame, after the embedding into a higher-dimensional space, is the orthogonal projection of an orthonormal basis. 
There exist several POVM-based definitions of coherence, and they are well-investigated.
Our aim is to identify properties specific to the particular case of tight frames, and to look for some applications.
All the POVM-based definitions use a Naimark extension. The frame-dependent coherence can be regarded as a particular case of POVM-based coherence, but it can be defined directly, without to use a Naimark extension. Its definition is a direct generalization 
of the basis-dependent $\ell_1$-norm of coherence, and  it offers a more accurate description because we can use a frame containing several orthogonal bases. A frame-invariant definition of coherence for qubits and multi-qubit systems is presented.
\end{abstract}

\begin{graphicalabstract}
\end{graphicalabstract}

\begin{highlights}
\item Frame-dependent coherence offers a more accurate description than the basis-dependent
\item The dependence  on the preferred frame is less strong than in case of preferred basis.
\item The used POVM-based definition becomes simpler in the particular case of frames.
\item For a multi-qubit system, a frame-independent definition of coherence can be given.
\end{highlights}


\end{frontmatter}




\section{Introduction}

Coherence is a fundamental property of quantum states arising from the superposition principle, 
but its  measurement is ambiguous. Estimation of coherence does not have a universal value.
In the basis-dependent case, the value of the coherence of a quantum state depends on
the basis in which it is measured.
This has been a point of concern, and there exist several attempts \cite{Baumgratz14,Chen16,Mandal20,Ma19,Radhakrishnan19,Hu17,Sperling18}  to remove this ambiguity.
A basis-independent definition has been obtained \cite{Ma19,Radhakrishnan19} by replacing the set of incoherent states 
by the set containing only the maximally mixed state. Other attempts use optimal bases \cite{Hu17}, quasiprobabilities \cite{Sperling18}
or POVMs \cite{Decker05,Bischof19}.

Our aim is to present in more details a definition of coherence, we call {\em frame-dependent coherence}, which can be regarded simultaneously either as a direct generalization of basis-dependent coherence or as a particular case for POVM-based coherence.
The use of frame-dependent coherence  is not an absolute way to quantify the coherence, but the dependence on the preferred 
frame is not as strong as in the case of basis-dependent coherence.

In a finite-dimensional Hilbert space, an {\em orthonormal basis} is 
 a system of orthogonal unit vectors satisfying the resolution of the identity, that is, such that the sum of the corresponding orthogonal projectors is the identity operator.
A {\em tight frame} is a system of vectors satisfying the resolution of the identity. In the case of a tight frame, it is not 
necessary the vectors to be of unit norm or orthogonal.
More than that, a frame can contain the null vector and  identical vectors.

Any vector of the Hilbert space can be represented as a linear
combination of the vectors of a tight frame, but generally, the representation is not unique.
Nevertheless, among all the possible representations, there
exists a standard one, defined by using the resolution of the identity satisfied by the vectors of the tight frame.
Similar to the case of an orthonormal basis, a matrix can be 
associated to a linear operator with respect to a tight 
frame.
The $\ell_1$-norm of coherence we use, similar to that used in  basis-dependent case \cite{Baumgratz14,Chen16},
is the sum of the modulus of the off-diagonal elements of the density matrix.

\section{Finite tight frames}

Let $(\mathcal{H},\langle.,.\rangle)$ be a $d$-dimensional complex Hilbert space, that is, a complex vector space $\mathcal{H}$ considered together with a scalar product 
$\mathcal{H}\!\times\!\mathcal{H}\!\rightarrow \!\mathbb{C}\!:\! (x,y)\!\mapsto\!\langle x,y\rangle$ satisfying the conditions
$\langle x,\alpha y\!+\!\beta z\rangle\!=\!\alpha \langle x,y\rangle\!+\!\beta\langle x,z\rangle$ and 
$\langle x,y\rangle\!=\!\overline{\langle y,x\rangle}$ for any 
$x,y,z\!\in\!\mathcal{H}$, \, $\alpha,\beta\!\in\!\mathbb{C}$, and the condition $\langle x,x\rangle\!>\!0$ for any $x\!\neq\!0$. The space $\mathcal{H}$ admits several equivalent representations:
\begin{itemize}
\item Standard representation
\begin{equation}
\mathcal{H}\!\equiv\!\mathbb{C}^d\!=\!\left\{ \, x\!=\!(x_1,x_2,...,x_d)\ |\ x_k\!\in\!\mathbb{C}\, \right\}, \qquad \langle x,y\rangle\!=\!\sum\limits_{k=1}^d\bar x_k\,y_k\, ;
\end{equation}
\item Dirac's representation as a space of column matrices
\begin{equation}
\mathcal{H}\!\equiv\!\left\{  \left. |x\rangle\!=\!\left(\!\!\begin{array}{c}
x_1\\[-1mm]
x_2\\[-1mm]
\vdots\\[-1mm]
x_d
\end{array}\!\!\right) \ \ \right|\  x_k\!\in\!\mathbb{C}\ \right\}\!,\quad 
\langle x,y\rangle\!\equiv\!\langle x|y\rangle\!=\! (\bar x_1\ \ \bar x_2\ ...\ \bar x_d)
\left(\!\!\begin{array}{c}
y_1\\[-1mm]
y_2\\[-1mm]
\vdots\\[-1mm]
y_d
\end{array}\!\!\right),
\end{equation}
where $\langle x|\!=\!(\bar x_1\ \ \bar x_2\ ...\ \bar x_d)$ is the ``bra'' matrix
corresponding to the ``ket'' $|x\rangle$;
\item Representation as a space of functions, defined on a set with $d$ elements
\begin{equation}
\mathcal{H}\!\equiv\!\left\{ \, \psi\!:\!\{ 1,2,...,d\}\rightarrow\!\mathbb{C}\, \right\}, \qquad \langle \varphi,\psi\rangle\!=\!\sum\limits_{k=1}^d\overline{\varphi(k)}\, \psi(k),
\end{equation}
or 
\begin{equation}
\mathcal{H}\!\equiv\!\left\{ \, \psi\!:\!\{ -s,-s\!+\!1,...,s\!-1,s\}\rightarrow\!\mathbb{C}\, \right\}, \qquad \langle \varphi,\psi\rangle\!=\!\sum\limits_{k=-s}^s\overline{\varphi(k)}\, \psi(k),
\end{equation}
in the case of odd $d\!=\!2s\!+\!1$.
\end{itemize}
Depending on the considered application, one representation may offer more formal advantages than the others. In each case, we  try to choose the most advantageous of them.
For example, for any $a,b\!\in\!\mathcal{H}$, the formula $A\!=\!|a\rangle\langle b|$ defines the linear operator
\begin{equation}
A\!:\!\mathcal{H}\rightarrow\!\mathcal{H}, \qquad A|x\rangle \!=\!|a\rangle\langle b|x\rangle
\end{equation}
described by the matrix
\begin{equation}
A\!=\!\left(\!\!\begin{array}{c}
a_1\\[-1mm]
a_2\\[-1mm]
\vdots\\[-1mm]
a_d
\end{array}\!\!\right) (\bar b_1\ \ \bar b_2\ ...\ \bar b_d)
\!=\!\left(\!\!\begin{array}{cccc}
a_1\bar b_1 & a_1\bar b_2 & \cdots & a_1\bar b_d\\
a_2\bar b_1 & a_2\bar b_2 & \cdots & a_2\bar b_d\\[-1mm]
\vdots & \vdots & \ddots & \vdots\\[-1mm]
a_d\bar b_1 & a_d\bar b_2 & \cdots & a_d\bar b_d
\end{array}\!\!\right).
\end{equation}

\noindent{\bf Definition 1.}  A set of vectors $\mathcal{B}\!=\!\{ |\psi_1\rangle, \, |\psi_2\rangle,\, ...\, , \, |\psi_d\rangle\}$, considered
in this order, is an {\em orthonormal basis} in $\mathcal{H}$ if the following two conditions are satisfied:
\begin{equation}
\langle\psi_j|\psi_k\rangle \!=\!\delta_{jk}\qquad \mbox{and}\qquad 
\sum\limits_{k=1}^d|\psi_k\rangle\langle \psi_k| \!=\!\mathbb{I},
\end{equation}
where $\mathbb{I}\!:\!\mathcal{H}\rightarrow\!\mathcal{H}, \ \mathbb{I}|\psi\rangle \!=\!|\psi\rangle$ is the identity operator, and 
\begin{equation}
\delta_{jk}\!=\!\left\{ \begin{array}{lll}
1 & \mbox{for} & j\!=\!k,\\
0 & \mbox{for} & j\!\neq\!k.
\end{array}\right.
\end{equation}
Any linear operator $A:\mathcal{H}\rightarrow\mathcal{H}$ admits the standard representation
\begin{equation}
\begin{array}{l}
A\equiv\mathbb{I}A\mathbb{I} \!=\!\sum\limits_{j,k=1}^d |\psi_j\rangle \langle \psi_j|A|\psi_k\rangle \langle \psi_k|\!=\!\sum\limits_{j,k=1}^d \langle \psi_j|A|\psi_k\rangle \, |\psi_j\rangle \langle \psi_k| .
\end{array}
\end{equation}
where
\begin{equation}
A\!=\!\left(\!\!\begin{array}{cccc}
\langle \psi_1|A|\psi_1\rangle  & \langle \psi_1|A|\psi_2\rangle  & \cdots & \langle \psi_1|A|\psi_d\rangle \\
\langle \psi_2|A|\psi_1\rangle  & \langle \psi_2|A|\psi_2\rangle  & \cdots & \langle \psi_2|A|\psi_d\rangle \\[-1mm]
\vdots & \vdots & \ddots & \vdots\\[-1mm]
\langle \psi_d|A|\psi_1\rangle  & \langle \psi_d|A|\psi_2\rangle  & \cdots & \langle \psi_d|A|\psi_d\rangle 
\end{array}\!\!\right).
\end{equation}
is the matrix of $A$ in the basis $\{ |\psi_1\rangle, \, |\psi_2\rangle,\, ...\, , \, |\psi_d\rangle\}$.\\[3mm]
\noindent{\bf Definition 2.} A set of vectors $\mathcal{F}\!=\!\{ |\varphi_1\rangle, \, |\varphi_2\rangle,\, ...\, , \, |\varphi_n\rangle\}$, considered
in this order, is a {\em tight frame} in $\mathcal{H}$ if the following single condition is satisfied \cite{Waldron18,Christensen03}:
\begin{equation}
\sum\limits_{k=1}^n|\varphi_k\rangle\langle \varphi_k| \!=\!\mathbb{I}.
\end{equation}
Any orthonormal basis is a tight frame. Generally, a tight frame contains more vectors than the dimension of the space, and the representation of a vector as a linear combination of the elements of such a tight frame is not unique.
Any element  $\psi\!\in\!\mathcal{H}$ admits the standard representation
\begin{equation}
|\psi\rangle \equiv\mathbb{I}|\psi\rangle \!=\!\sum\limits_{k=1}^n |\varphi_k\rangle \langle\varphi_k|\psi\rangle \!=\!\sum\limits_{k=1}^n \langle \varphi_k|\psi\rangle \, |\varphi_k\rangle 
\end{equation}
as a linear combination of $|\varphi_1\rangle, \, |\varphi_2\rangle,\, ...\, , \, |\varphi_n\rangle$.
This representation  is a privileged one: for any $\alpha_k\!\in\!\mathbb{C}$ such that 
\begin{equation}
|\psi\rangle \!=\!\sum\limits_{k=1}^n  \alpha_k\, |\varphi_k\rangle 
\qquad \mbox{we have}\qquad 
\sum\limits_{k=1}^n  |\alpha_k|^2\geq \sum\limits_{k=1}^n |\langle \varphi_k|\psi\rangle |^2.
\end{equation}
For example, in the case $\mathcal{H}\!=\!\mathbb{C}^2$, the vectors
{\small
\begin{equation}
|\varphi_0\rangle\!=\!\left( \!\!
\begin{array}{c}
\sqrt{\frac{2}{3}}\\[2mm]
0
\end{array} \!\!\right),\quad |\varphi_1\rangle\!=\!\left(  \!\!
\begin{array}{r}
-\frac{1}{\sqrt{6}}\\[2mm]
\frac{1}{\sqrt{2}}
\end{array} \!\!\right),\quad |\varphi_2\rangle\!=\!\left(  \!\!
\begin{array}{r}
-\frac{1}{\sqrt{6}}\\[2mm]
-\frac{1}{\sqrt{2}}
\end{array} \!\!\right),
\end{equation}
}
\noindent satisfying the relation $|\varphi_0\rangle\!+\!|\varphi_1\rangle\!+\!|\varphi_2\rangle\!=\!0$, form a tight frame in $\mathbb{C}^2$,
\begin{equation}
|\varphi_0\rangle\langle \varphi_0|\!+\!|\varphi_1\rangle\langle \varphi_1|\!+\!|\varphi_2\rangle\langle \varphi_2|\!=\!
{\small
\left( \!\!\begin{array}{cc}
1 & 0\\
0 & 1
\end{array} \!\!\right)},
\end{equation}
and for any $|\psi\rangle\!\in \!\mathbb{C}^2$ and any $\lambda\!\in \!\mathbb{C}$, we have
\begin{equation}
\begin{array}{l}
|\psi\rangle \!=\!\sum\limits_{k=0}^2 |\varphi_k\rangle \langle \varphi_k|\psi\rangle \!=\!\sum\limits_{k=0}^2 |\varphi_k\rangle (\langle \varphi_k|\psi\rangle \!+\!\lambda),\\
\sum\limits_{k=0}^2 |\langle \varphi_k|\psi\rangle \!+\!\lambda|^2\!=\!\sum\limits_{k=0}^2 |\langle \varphi_k|\psi\rangle|^2 \!+\!|\lambda|^2\!\geq\!\sum\limits_{k=0}^2 |\langle \varphi_k|\psi\rangle|^2.
\end{array}.
\end{equation}
A frame can contain the null vector or identical vectors. For example,
{\small
\begin{equation}
\begin{array}{l}
\left\{ \left(\!\!
\begin{array}{c}
1\\[1mm]
0
\end{array}\!\!\right)\!, \left( \!\!
\begin{array}{c}
0\\[1mm]
1
\end{array}\!\!\right)\!, \left( \!\!
\begin{array}{c}
0\\[1mm]
0
\end{array}\!\!\right)\right\},\ \  
\left\{ \left( \!\!
\begin{array}{c}
1\\[1mm]
0
\end{array}\!\!\right)\!, \left( \!\!
\begin{array}{c}
0\\[1mm]
1
\end{array}\!\!\right)\!, \left( \!\!
\begin{array}{c}
0\\[1mm]
0
\end{array}\!\!\right)\!, \left( \!\!
\begin{array}{c}
0\\[1mm]
0
\end{array}\!\!\right)\right\},\ 
 \left\{ \left( \!\!
\begin{array}{c}
\frac{1}{\sqrt{2}}\\[1mm]
0
\end{array}\!\!\right)\!, \left( \!\!
\begin{array}{c}
\frac{1}{\sqrt{2}}\\[1mm]
0
\end{array}\!\!\right)\!, \left( \!\!
\begin{array}{c}
0\\[1mm]
\frac{1}{\sqrt{2}}
\end{array}\!\!\right)\!, \left( \!\!
\begin{array}{c}
0\\[1mm]
\frac{1}{\sqrt{2}}
\end{array}\!\!\right)\right\}
\end{array}
\end{equation}
}
\noindent are tight frames in $\mathbb{C}^2$.
Any linear operator $A\!:\!\mathcal{H}\!\rightarrow\!\mathcal{H}$ admits the standard representation
\begin{equation}
\begin{array}{l}
A\equiv\mathbb{I}A\mathbb{I} \!=\!\sum\limits_{j,k=1}^n |\varphi_j\rangle \langle \varphi_j|A|\varphi_k\rangle \langle \varphi_k|\!=\!\sum\limits_{j,k=1}^n \langle \varphi_j|A|\varphi_k\rangle \, |\varphi_j\rangle \langle \varphi_k| .
\end{array}
\end{equation}
where
\begin{equation}
A\!=\!\left(\!\!\begin{array}{cccc}
\langle \varphi_1|A|\varphi_1\rangle  & \langle \varphi_1|A|\varphi_2\rangle  & \cdots & \langle \varphi_1|A|\varphi_n\rangle \\
\langle \varphi_2|A|\varphi_1\rangle  & \langle \varphi_2|A|\varphi_2\rangle  & \cdots & \langle \varphi_2|A|\varphi_n\rangle \\[-1mm]
\vdots & \vdots & \ddots & \vdots\\[-1mm]
\langle \varphi_n|A|\varphi_1\rangle  & \langle \varphi_n|A|\varphi_2\rangle  & \cdots & \langle \varphi_n|A|\varphi_n\rangle 
\end{array}\!\!\right).
\end{equation}
is the matrix of $A$ in the frame $\{ |\varphi_1\rangle, \, |\varphi_2\rangle,\, ...\, , \, |\varphi_n\rangle\}$.
If $A$ is Hermitian, $A\!=\!A^\dag$, then
\begin{equation}
\langle \varphi_j|A|\varphi_k\rangle \!\equiv\!\langle \varphi_j,A\varphi_k\rangle \!=\!\langle A\varphi_j,\varphi_k\rangle \!=\!\overline{\langle \varphi_k,A\varphi_j\rangle} \!\equiv\!\overline{\langle \varphi_k|A|\varphi_j\rangle}.
\end{equation}
As concerne the trace of $A$, we have
\begin{equation}
\mathrm{tr}\, A\!=\!\mathrm{tr}(\mathbb{I} A)\!=\!\mathrm{tr}(\sum\limits_{k=1}^n|\varphi_k\rangle\langle \varphi_k| A)\!=\!
\sum\limits_{k=1}^n\mathrm{tr}(|\varphi_k\rangle\langle \varphi_k| A)\!=\!\sum\limits_{k=1}^n\langle \varphi_k| A|\varphi_k\rangle.
\end{equation}

If $\mathcal{F}\!=\!\{\,  |\varphi_1\rangle, \, |\varphi_2\rangle,\, ...\, , \, |\varphi_n\rangle\}$ is a tight frame and  
$U\!:\!\mathcal{H}\!\rightarrow\!\mathcal{H}$ a unitary transform, then
\begin{equation}
\sum\limits_{k=1}^n U|\varphi_k\rangle\langle \varphi_k|U^\dag \!=\!U\sum\limits_{k=1}^n |\varphi_k\rangle\langle \varphi_k|U^\dag \!=\!U\mathbb{I}U^\dag\!=\!\mathbb{I},
\end{equation}
that is, $U\!\mathcal{F}\!=\!\{ U|\varphi_1\rangle, \, U|\varphi_2\rangle,\, ...\, , \, U|\varphi_n\rangle\}$
is also a tight frame.


\section{Frame-dependent coherence of a quantum state}

A pure state of a quantum system with  Hilbert space $\mathcal{H}$ (we consider only the finite-dimensional case), is described by a normed element $\psi\!\in\!\mathcal{H}$. The corresponding orthogonal projector $\varrho_\psi\!=\!|\psi\rangle\langle \psi|$ is self-adjoint (also called Hermitian),
$\varrho_\psi\!\geq\!0$ (its eigenvalues 1 and  0 are non-negative) and $\mathrm{tr}\, \varrho_\psi \!=\!1$ (sum of eigenvalues is 1).

The self-adjoint operators $\varrho\!:\!\mathcal{H}\!\rightarrow\!\mathcal{H}$ satisfying the conditions $\varrho\!\geq\!0$
(all eigenvalues are non-negative) and $\mathrm{tr}\, \varrho \!=\!1$ (sum of eigenvalues is $1$) are called 
{\it density operators} and they describe generalized states of the quantum system.
For any density operator $\varrho$, in the Hilbert space $\mathcal{H}$, there exists an orthonormal basis 
$\mathfrak{B}\!=\!\{ |\eta_1\rangle, \, |\eta_2\rangle,\, ...\, , \, |\eta_d\rangle\}$ containing only eigenvectors of $\varrho$
and corresponding to non-negative eigenvalues $\lambda_1,\, \lambda_2,....,\lambda_d$ with
$\sum_{k=1}^d\lambda_k\!=\!1$. Thus, $\varrho$ admits the spectral resolution
\begin{equation}
\varrho\!=\!\sum\limits_{k=1}^d \lambda_k |\eta_k\rangle\langle \eta_k|,
\end{equation}
that is, $\varrho$ is a statistical mixture of the pure states $\eta_1, \, \eta_2,\, ...\, , \, \eta_d$ with the probabilities
$\lambda_1,\, \lambda_2,....,\lambda_d$.\\[3mm]
\noindent{\bf Definition 3.} Let  $\mathfrak{B}\!=\!\{ |\psi_1\rangle, \, |\psi_2\rangle,\, ...\, , \, |\psi_d\rangle\}$ be an  orthonormal basis of $\mathcal{H}$. The basis-dependent $\ell_1$-{\em norm of coherence} related to $\mathfrak{B}$ of a quantum state $\varrho\!:\!\mathcal{H}\!\rightarrow \!\mathcal{H}$ is \cite{Baumgratz14,Chen16}
\begin{equation}
\mathcal{C}_\mathfrak{B}(\varrho)\!=\!\sum\limits_{
j\neq k
}|\langle \psi_j|\varrho|\psi_k\rangle|\!=\!2\sum\limits_{
j< k
}|\langle \psi_j|\varrho|\psi_k\rangle|.
\end{equation}
\noindent{\bf Definition 4.} Let  $\mathfrak{F}\!=\!\{ |\varphi_1\rangle, \, |\varphi_2\rangle,\, ...\, , \, |\varphi_n\rangle\}$ be a tight frame of $\mathcal{H}$. The {\em frame-dependent coherence} related to $\mathfrak{F}$ of a quantum state $\varrho\!:\!\mathcal{H}\!\rightarrow \!\mathcal{H}$ is
\begin{equation}
\mathcal{C}_\mathfrak{F}(\varrho)\!=\!\mbox{$\frac{d}{n}$}\sum\limits_{
j\neq k
}|\langle \varphi_j|\varrho|\varphi_k\rangle|\!=\!2\mbox{$\frac{d}{n}$}\sum\limits_{
j=1}^{n-1}\sum\limits_{
k=j+1}^{n}|\langle \varphi_j|\varrho|\varphi_k\rangle|.
\end{equation}
Frame-dependent coherence is a natural extension for basis-dependent coherence. If the used tight frame is an orthonormal basis, 
then the corresponding frame-dependent coherence coincides with the usual basis-dependent  $\ell_1$- norm of coherence.

Similar to the case of an orthonormal basis \cite{Mandal20}, the matrix elements $\langle \varphi_j|\varrho|\varphi_k\rangle $ 
of a density operator $\varrho$ can be expressed in terms of the mean values of the observables 
\begin{equation}
W_{\!jk}\!=\!\left\{ \begin{array}{lll}
| \varphi_j\rangle\langle  \varphi_j| & \mathrm{if} & j\!=\!k\\[1mm]
\frac{1}{2}(| \varphi_j\rangle\langle  \varphi_k|\!+\!| \varphi_k\rangle\langle  \varphi_j|) & \mathrm{if} & j\!>\!k\\[1mm]
\frac{\mathrm{i}}{2}(| \varphi_j\rangle\langle  \varphi_k|\!-\!| \varphi_k\rangle\langle  \varphi_j|) & \mathrm{if} & j\!<\!k.
\end{array}\right.
\end{equation}
For example, for $j\!<\!k$, we have 
\begin{equation}
\langle  \varphi_j|\varrho| \varphi_k\rangle\!=\!\mathrm{tr}(\varrho\ | \varphi_k\rangle\langle  \varphi_j)|\!=\!\langle W_{\!kj}\rangle_\varrho\!+\!\mathrm{i}\langle W_{\!jk}\rangle_\varrho.
\end{equation}
and consequently
\begin{equation}
\mathcal{C}_\mathfrak{F}(\varrho)\!=\!2\mbox{$\frac{d}{n}$}\sum\limits_{
j=1}^{n-1}\sum\limits_{
k=j+1}^{n}|\langle W_{\!kj}\rangle_\varrho\!+\!\mathrm{i}\langle W_{\!jk}\rangle_\varrho|.
\end{equation}
Generally, in the case of a frame, the elements $|\varphi_k\rangle$ and hence the operators $W_{\!jk}$ are not independent. Consequently, in order to obtain the frame-dependent coherence of $\varrho$, we have to measure the mean values for only $d^2$ of the $n^2$ operators  $W_{\!jk}$. In addition, we have the possibility to choose the most experimentally accessible.

\noindent{\bf Theorem 1.} {\em Frame-dependent coherence has the properties:
\begin{itemize}
\item Non-negativity: \ \ $\mathcal{C}_\mathfrak{F}(\varrho)\!\geq \!0$;
\item Convexity:\ \  $\mathcal{C}_\mathfrak{F}(\sum\limits_{m=1}^\ell p_m\, \varrho_m)\!\leq \!\sum\limits_{m=1}^\ell p_m\, \mathcal{C}_\mathfrak{F}(\varrho_m)$;
\item Unitarily-invariant: \ \  $\mathcal{C}_{U\mathfrak{F}}(U\varrho U^\dag)\!=\!\mathcal{C}_\mathfrak{F}(\varrho)$;
\end{itemize}
where $p_m\geq 0$ are such that $\sum\limits_{m=1}^\ell p_m\!=\!1$, and $U\!:\!\mathcal{H}\!\rightarrow\!\mathcal{H}$ is an arbitrary unitary operator.}\\[3mm]
{\it Proof.} We have 
\[
\begin{array}{l}
\mathcal{C}_\mathfrak{F}(\sum\limits_{m=1}^\ell p_m\, \varrho_m)\!=\!\frac{d}{n}\sum\limits_{
j\neq k
}|\langle \varphi_j|\sum\limits_{m=1}^\ell p_m\, \varrho_m|\varphi_k\rangle|\!=\!\frac{d}{n}\sum\limits_{
j\neq k
}|\sum\limits_{m=1}^\ell p_m\langle \varphi_j|\varrho_m|\varphi_k\rangle|\\
\leq\! \frac{d}{n}\sum\limits_{
j\neq k }\sum\limits_{m=1}^\ell p_m|\langle \varphi_j|\varrho_m|\varphi_k\rangle|\!=\!
\sum\limits_{m=1}^\ell p_m\frac{d}{n}\sum\limits_{
j\neq k }|\langle \varphi_j|\varrho_m|\varphi_k\rangle|\!=\!\sum\limits_{m=1}^\ell p_m\, \mathcal{C}_\mathfrak{F}(\varrho_m),
\end{array}
\]
and
\[
\mathcal{C}_{U\mathfrak{F}}(U\varrho U^\dag)\!=\!\mbox{$\frac{d}{n}$}\sum\limits_{
j\neq k
}|\langle \varphi_j|U^\dag U\varrho U^\dag U|\varphi_k\rangle|\!=\!\mbox{$\frac{d}{n}$}\sum\limits_{
j\neq k
}|\langle \varphi_j|\varrho |\varphi_k\rangle|\!=\!\mathcal{C}_\mathfrak{F}(\varrho).
\]

\section{Some examples}

\subsection{Coherence of some qubit quantum states}

In Dirac notation, the qubit Hilbert space is 
\begin{equation}
\mathbb{C}^2\!=\!\left\{ \left. |\psi\rangle\!=\!\left(\!\!\begin{array}{c}
\alpha\\
\beta
\end{array}\!\!\right)\ \right|\ \alpha,\beta\!\in\!\mathbb{C}\ \right\},
\end{equation}
usually described by using the canonical (computational) basis
\begin{equation}
\left\{ \ |0\rangle\!=\!\left(\!\!\begin{array}{c}
1\\
0
\end{array}\!\!\right),\ |1\rangle\!=\!\left(\!\!\begin{array}{c}
0\\
1
\end{array}\!\!\right)\ \right\}.
\end{equation}
a) The coherence of the pure quantum state
\begin{equation}
\varrho_0\!=\!|0\rangle\langle 0|\!=\!
{\small
\left( \!\!\begin{array}{rr}
1 &  0 \\[1mm]
0 & 0
\end{array} \!\!\right) }
\end{equation}
with respect to the orthonormal basis 
\begin{equation}
\mathfrak{B}_\lambda\!=\!
{\small
\left\{ \ |\psi_1\rangle\!=\!\left(\!\!\!
\begin{array}{c}
\cos\lambda\\
\sin \lambda
\end{array}\!\!\!\right)\!,\ |\psi_2\rangle\!=\!\left( \!\!\!
\begin{array}{r}
-\sin \lambda\\
\cos \lambda
\end{array}\!\!\!\right)\ \right\},\!
}
\end{equation}
depending on a parameter $\lambda\!\in\![0,2\pi)$,  is 
\begin{equation}
\mathcal{C}_{\mathfrak{B}_\lambda}(\varrho_0)\!=\!2|\langle \psi_1|\varrho_0|\psi_2\rangle|\!=\!\frac{1}{2}|\sin 2\lambda|, 
\end{equation}
and the coherence with respect to the regular polygonal frame
\begin{equation}\label{polyframe}
{\small
\mathfrak{F}_n\!=\!\left\{ \left.|\varphi_k\rangle\!=\!
 \sqrt{\frac{2}{n}}\left(  \!\!
\begin{array}{c}
\cos\frac{2k\pi}{n}\\[1mm]
\sin\frac{2k\pi}{n}
\end{array} \!\!\right)\ \right| \ \ k\!\in\!\{0,1,2,...,n\!-\!1\}\ \right\}
}
\end{equation}
where $n\!\in\!\{ 3,4,5,...\}$, is (see Fig.1a)
\begin{equation}
\mathcal{C}_{\mathfrak{F}_n}(\varrho_0)\!=\!\frac{2}{n}\sum\limits_{j\neq k}|\langle \varphi_j|\varrho_0 |\varphi_k\rangle|\!=\!\frac{4}{n^2}\sum\limits_{j\neq k}\left|\cos\frac{2j\pi}{n}\, \cos \frac{2k\pi}{n}\right|.
\end{equation}
 b) The coherence of the maximally mixed state
\begin{equation}
\varrho_1\!=\!
{\small
\left( \!\!\begin{array}{rr}
\frac{1}{2} &  0 \\[1mm]
0 & \frac{1}{2}
\end{array} \!\!\right) }
\end{equation}
with respect to the orthonormal basis $\mathfrak{B}_\lambda$ is
\begin{equation}
\mathcal{C}_{\mathfrak{B}_\lambda}(\varrho_1)\!=\!2|\langle \psi_1|\varrho_1|\psi_2\rangle|\!=\!0, 
\end{equation}
and the coherence with respect to the regular polygonal frame $\mathfrak{F}_n$ is (see Fig.1b)
\begin{equation}
\mathcal{C}_{\mathfrak{F}_n}(\varrho_1)\!=\!\frac{2}{n}\sum\limits_{j\neq k}|\langle \varphi_j|\varrho_1 |\varphi_k\rangle|\!=\!\frac{2}{n^2}\sum\limits_{j\neq k}\left|\cos\frac{2(j\!-\!k)\pi}{n}\right|.
\end{equation}
 c) The coherence of the quantum state
\begin{equation}
\varrho_2\!=\!
{\small
\left( \!\!\begin{array}{rr}
\frac{1}{4} &  0 \\[1mm]
0 & \frac{3}{4}
\end{array} \!\!\right) }
\end{equation}
with respect to the orthonormal basis $\mathfrak{B}_\lambda$ is
\begin{equation}
\mathcal{C}_{\mathfrak{B}_\lambda}(\varrho_2)\!=\!2|\langle \psi_1|\varrho_2|\psi_2\rangle|\!=\!\frac{1}{2}|\sin 2\lambda|, 
\end{equation}
and the coherence with respect to the regular polygonal frame $\mathfrak{F}_n$ is (see Fig.1c)
\begin{equation}
\mathcal{C}_{\mathfrak{F}_n}(\varrho_2)\!=\!\frac{1}{n^2}\sum\limits_{j\neq k}\left|\cos\frac{2j\pi}{n}\, \cos \frac{2k\pi}{n}\!+\!3\sin\frac{2j\pi}{n}\, \sin \frac{2k\pi}{n}\right|.
\end{equation}
 d) The coherence of the quantum state
\begin{equation}
\varrho_3\!=\!
{\small
\left( \!\!\begin{array}{rr}
\frac{1}{2} &  -\frac{1}{4} \\[1mm]
-\frac{1}{4} & \frac{1}{2}
\end{array} \!\!\right) }
\end{equation}
with respect to the orthonormal basis $\mathfrak{B}_\lambda$ is
\begin{equation}
\mathcal{C}_{\mathfrak{B}_\lambda}(\varrho_3)\!=\!2|\langle \psi_1|\varrho_3|\psi_2\rangle|\!=\!\frac{1}{2}|\cos 2\lambda|, 
\end{equation}
and the coherence with respect to the regular polygonal frame $\mathfrak{F}_n$ is (see Fig.1d)
\begin{equation}
\mathcal{C}_{\mathfrak{F}_n}(\varrho_3)\!=\!\frac{2}{n}\sum\limits_{j\neq k}|\langle \varphi_j|\varrho_3 |\varphi_k\rangle|\!=\!\frac{1}{n^2}\sum\limits_{j\neq k}\left|2\cos\frac{2(j\!-\!k)\pi}{n}\!-\! \sin \frac{2(j\!+\!k)\pi}{n}\right|.
\end{equation}
One can remark that, for large $n$, the frame coherence  tends to be independent on $n$, and can be regarded as a kind of frame-independent coherence.

\begin{figure}[h]
a)\includegraphics[scale=1.0]{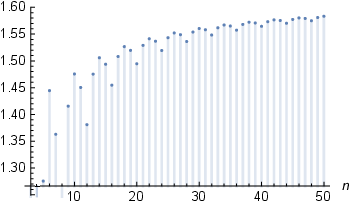}\qquad 
b)\includegraphics[scale=1.0]{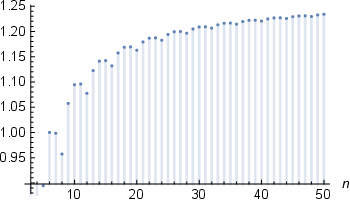}\\[5mm]
c)\includegraphics[scale=1.0]{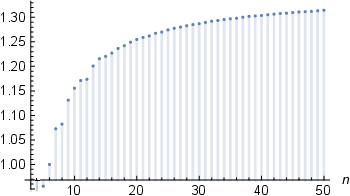}\qquad 
d)\includegraphics[scale=1.0]{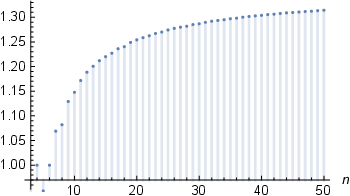}
\caption{Frame coherence: a) $\mathcal{C}_{\mathfrak{F}_n}(\varrho_0)$, \  b) $\mathcal{C}_{\mathfrak{F}_n}(\varrho_1)$,
\ c) $\mathcal{C}_{\mathfrak{F}_n}(\varrho_2)$, \ d) $\mathcal{C}_{\mathfrak{F}_n}(\varrho_3)$
for $n\!\in\!\{ 3,4,5,...,50\}$. }
\end{figure}

\noindent e) The coherence of any quantum state $\varrho$ with respect to the frame 
\begin{equation}\label{triangular}
\mathfrak{F}_3\!=\!
{\small
\left\{ 
|\varphi_0\rangle\!=\!\sqrt{\frac{2}{3}}\left( \!\!
\begin{array}{c}
1\\[1mm]
0
\end{array} \!\!\right)\!,\, |\varphi_1\rangle\!=\!\sqrt{\frac{2}{3}}\left(  \!\!
\begin{array}{c}
-\frac{1}{2}\\[1mm]
\frac{\sqrt{3}}{2}
\end{array} \!\!\right)\!,\, |\varphi_2\rangle\!=\!\sqrt{\frac{2}{3}}\left(  \!\!
\begin{array}{c}
-\frac{1}{2}\\[1mm]
-\frac{\sqrt{3}}{2}
\end{array} \!\!\right)
\right\};}
\end{equation}
is non-null. Indeed, $\mathcal{C}_{\mathfrak{F}_3}(\varrho)\!=\!\frac{4}{3}(|\langle \varphi_0|\varrho |\varphi_1\rangle|\!+\!|\langle \varphi_0|\varrho| \varphi_2\rangle|\!+\!|\langle  \varphi_1|\varrho |\varphi_2\rangle|)$ and
\[
\mathcal{C}_{\mathfrak{F}_3}(\varrho)\!=\!0\ \Leftrightarrow\ \left\{
\begin{array}{l}
0\!=\!|\langle \varphi_0|\varrho |\varphi_1\rangle|\\
0\!=\!|\langle \varphi_0|\varrho |\varphi_2\rangle|\!=\!|\langle \varphi_0|\varrho |\varphi_0\rangle|\!+\!|\langle \varphi_0|\varrho |\varphi_1\rangle|\\
0\!=\!|\langle \varphi_1|\varrho |\varphi_2\rangle|\!=\!|\langle \varphi_1|\varrho |\varphi_0\rangle|\!+\!|\langle \varphi_1|\varrho |\varphi_1\rangle|.
\end{array}\right.\ \Leftrightarrow\ \varrho\!=\!0.
\]
The matrix of an arbitrary quantum state, in the canonical basis, can be represented as 
\begin{equation}\label{qubitstate}
\varrho\!=\!
\left( \!\!\begin{array}{ll}
a & b\, \mathrm{e}^{\mathrm{i}\theta}\\
 b\, \mathrm{e}^{-\mathrm{i}\theta} & 1\!-\!a 
\end{array} \!\!\right),\quad \mbox{where}\quad 
\begin{array}{c}
0\!\leq\!a\!\leq\!1,\ \ \ 0\!\leq\!\theta\!<\!2\pi\\[1mm]
-\sqrt{a(1\!-\!a)}\!\leq\!b\!\leq\!\sqrt{a(1\!-\!a)}.
\end{array}
\end{equation}
Figure 2 presents the coherence $\mathcal{C}_{\mathfrak{F}_3}(\varrho)$ in the particular cases $\theta\!=\!0$ and $\theta\!=\!\frac{\pi}{3}$.

\begin{figure}[t]
\includegraphics[scale=1.1]{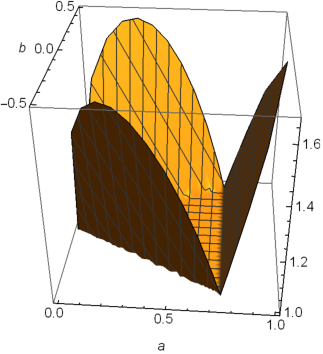}\quad 
\includegraphics[scale=1.2]{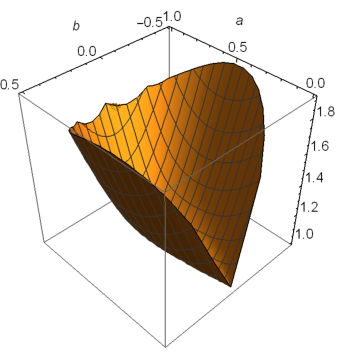}
\caption{Frame coherence: $\mathcal{C}_{\mathfrak{F}_3}(\varrho)$ of the state (\ref{qubitstate}) in the cases $\theta\!=\!0$ and $\theta\!=\!\frac{\pi}{3}$. }
\end{figure}

\subsection{Coherence of a qutrit quantum state}

In the case of a qutrit described by the Hilbert space $\mathbb{C}^3$, 
the coherence of the quantum state
\begin{equation}\label{rho3}
\varrho\!=\!
{\small
\frac{1}{6}\left( \!\!\begin{array}{ccc}
1 & 0 & 0\\
0 & 2 & 0\\
0 & 0 & 3
\end{array} \!\!\right)}
\end{equation}
is:
\begin{equation}
\mathcal{C}_{\mathfrak{B}_{can}}(\varrho)\!=\!0\ \ \mbox{if we choose the basis}\ \  
\mathfrak{B}_{can}\!=\!
{\small
\left\{ |\psi_1\rangle\!=\!\left(\!\!
\begin{array}{c}
1\\
0\\
0
\end{array}\!\!\right)\!, |\psi_2\rangle\!=\!\left( \!\!
\begin{array}{c}
0\\
1\\
0
\end{array}\!\!\right)\!,|\psi_3\rangle\!=\!\left( \!\!
\begin{array}{c}
0\\
0\\
1
\end{array}\!\!\right)
\right\}\!;}
\end{equation}
$\mathcal{C}_{\mathfrak{B}_{can}^*}(\varrho)\!=\!0.577 \ \ \mbox{if we choose the corresponding complementary basis}$\\
\begin{equation}
\mathfrak{B}_{can}^{*}\!=\!
{\small
\left\{ \!
F^\dag |\psi_1\rangle\!=\!\frac{1}{\sqrt{3}}\!\left( \!\!
\begin{array}{c}
1\\
1\\
1
\end{array}\!\!\right)\!,\, F^\dag |\psi_2\rangle\!=\!\frac{1}{\sqrt{3}}\!\left( \!\!
\begin{array}{l}
1\\
\mathrm{e}^{ \frac{2\pi \mathrm{i}}{3}}\\
\mathrm{e}^{ -\frac{2\pi \mathrm{i}}{3}}
\end{array}\!\!\!\right)\!,\, F^\dag|\psi_3\rangle\!=\!\frac{1}{\sqrt{3}}\!\left( \!\!
\begin{array}{l}
1\\
\mathrm{e}^{ -\frac{2\pi \mathrm{i}}{3}}\\
\mathrm{e}^{ \frac{2\pi \mathrm{i}}{3}}
\end{array}\!\!\!\right)\!
\right\}\!;}
\end{equation}
$\mathcal{C}_{\mathfrak{F}}(\varrho)\!=\!1.010 \ \ \mbox{if we choose the frame containing both $\mathfrak{B}_3$ and $\mathfrak{B}_3^*$}$,
\begin{equation}
\mathfrak{F}\!=\!
{\small
\left\{ 
\frac{1}{\sqrt{2}}|\psi_1\rangle,\, \frac{1}{\sqrt{2}}|\psi_2\rangle,\, \frac{1}{\sqrt{2}}|\psi_3\rangle,\,
\frac{1}{\sqrt{2}}F^\dag |\psi_1\rangle,\, \frac{1}{\sqrt{2}}F^\dag |\psi_2\rangle,\, \frac{1}{\sqrt{2}}F^\dag|\psi_3\rangle
\right\}\!;}
\end{equation}
$\mathcal{C}_{\mathfrak{F}_{tetra}}(\varrho)\!=\!0.75 \ \ \mbox{if we choose the tetrahedral frame \cite{Cotfas10,Cotfas24} }$\\
\begin{equation}
\mathfrak{F}_{tetra}\!=\!
{\small
  \left\{\frac{1}{2}\!\left(\!\!
\begin{array}{r}
-1\\
1\\
1
\end{array}\!\!
\right)\! , \frac{1}{2}\!\left(\!\!
\begin{array}{r}
1\\
-1\\
1
\end{array}\!\!
\right) \!,\frac{1}{2}\!\left(\!\!
\begin{array}{r}
1\\
1\\
-1
\end{array}\!\!
\right)\!, \frac{1}{2}\!\left(\!\!
\begin{array}{r}
-1\\
-1\\
-1
\end{array}\!\!
\right)
\right\};
}
\end{equation}
$\mathcal{C}_{\mathfrak{F}_{ico}}(\varrho)\!=\!1.135 \ \ \mbox{if we choose the icosahedral frame (regular icosahedron) \cite{Cotfas10,Cotfas24}}$\\
\begin{equation}{\mathfrak{F}_{ico}}\!=\!
{\small
\left\{ 
\frac{1}{\eta}\!\left(\!\!\!
\begin{array}{r}
1\\
\tau\\
0
\end{array}\!\!\!
\right) , \frac{1}{\eta}\!\left(\!\!\!
\begin{array}{r}
-1\\
\tau\\
0
\end{array}\!\!\!
\right) , 
\frac{1}{\eta}\!\left(\!\!\!
\begin{array}{r}
- \tau\\
0\\
1
\end{array}\!\!\!
\right),  
\frac{1}{\eta}\!\left(\!\!\!
\begin{array}{r}
0\\
-1\\
\tau
\end{array}\!\!\!
\right) , 
\frac{1}{\eta}\!\left(\!\!\!
\begin{array}{r}
\tau\\
0\\
1
\end{array}\!\!\!
\right) , \frac{1}{\eta}\!\left(\!\!\!
\begin{array}{r}
0\\
1\\
\tau
\end{array}\!\!\!
\right)
\right\},
}
\end{equation}
where $\tau\!=\!\frac{1\!+\!\sqrt{5}}{2}$, \ $\eta\!=\!\sqrt{5\!+\!\sqrt{5}}$, \ $F\!:\!\mathbb{C}^3\!\rightarrow \!\mathbb{C}^3$,\\
\begin{equation}
F\!=\!
{\small
\frac{1}{\sqrt{3}}\left( \!\!\begin{array}{lll}
1 & 1 & 1\\[1mm]
1 & \mathrm{e}^{- \frac{2\pi \mathrm{i}}{3}}& \mathrm{e}^{ \frac{2\pi \mathrm{i}}{3}}\\[1mm]
1 & \mathrm{e}^{ \frac{2\pi \mathrm{i}}{3}} & \mathrm{e}^{- \frac{2\pi \mathrm{i}}{3}}
\end{array} \!\!\right)}
\end{equation}
is the Fourier transform, and $F^\dag$  its adjoint. Larger icosahedral frames can be obtained by including vectors corresponding to the vertices of certain dodecahedrons, icosidodecahedrons and  by using unitary transformations.


\section{Variation of the basis-dependent coherence from  one basis to another one}

If  $\mathfrak{B}\!=\!\{ |\psi_1\rangle, \, |\psi_2\rangle,\, ...\, , \, |\psi_d\rangle\}$ and 
$\mathfrak{B}'\!=\!\{ |\varphi_1\rangle, \, |\varphi_2\rangle,\, ...\, , \, |\varphi_d\rangle\}$ are two 
orthonormal bases in $\mathcal{H}$, then the tight frame 
\begin{equation}
\mathfrak{F}(t)\!=\!\{ \sqrt{1\!-\!t}|\psi_1\rangle, \, \sqrt{1\!-\!t}|\psi_2\rangle,\, ...\, , \, \sqrt{1\!-\!t}|\psi_d\rangle, 
\sqrt{t}|\varphi_1\rangle, \, \sqrt{t}|\varphi_2\rangle,\, ...\, , \, \sqrt{t}|\varphi_d \rangle\}
\end{equation}
depending on  $t \!\in\! [0,1]$ can be regarded as a continuous deformation of  $\mathfrak{B}$ to $\mathfrak{B}'$. We have
\begin{equation}
\mathcal{C}_{\mathfrak{F}(t)}(\varrho)\!=\!\frac{1}{2}(1\!-\!t)\sum\limits_{j< k}|\langle \psi_j|\varrho|\psi_k\rangle|\!+\!
\frac{1}{2}\sqrt{t(1\!-\!t)}\sum\limits_{j=1}^d\sum\limits_{k=1}^d|\langle \psi_j|\varrho|\varphi_k\rangle|\!+
\!\frac{1}{2}t\sum\limits_{j< k}|\langle \varphi_j|\varrho|\varphi_k\rangle| ,
\end{equation}
and one can remark that 
\begin{equation}
\mathfrak{F}(0)\!=\!\{|\psi_1\rangle, \, |\psi_2\rangle,\, ...\, , \, |\psi_d\rangle, 
0,\, 0,\, ...\, ,\, 0 \}\quad \Rightarrow\quad\mathcal{C}_{\mathfrak{B}}(\varrho) \!=\!2\,\mathcal{C}_{\mathfrak{F}(0)}(\varrho) ,
\end{equation}
\begin{equation}
\mathfrak{F}(1)\!=\!\{0,\, 0,\, ...\,,\,0,\, |\varphi_1\rangle, \, |\varphi_2\rangle,\, ...\, , \, |\varphi_d\rangle 
\}\quad \Rightarrow\quad \mathcal{C}_{\mathfrak{B}'}(\varrho) \!=\!2\,\mathcal{C}_{\mathfrak{F}(1)}(\varrho),
\end{equation}
for any quantum state $\varrho$.

So, we can connect two orthonormal bases by a family of tight frames and investigate explicitly how the coherence
changes when we go from one orthonormal basis to another.
In Fig. 3, we describe how the coherence of the state (\ref{rho3}) changes when we pass from the  canonical basis  
$\mathfrak{B}_{can}$ to the complementary basis $\mathfrak{B}_{can}^*$.

\begin{figure}[t]
\includegraphics[scale=1.2]{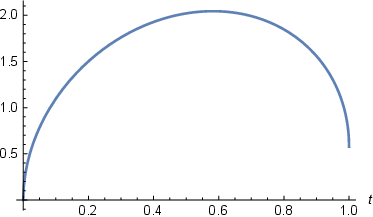}
\caption{Variation of the basis-dependent coherence from  $\mathfrak{B}$ to $\mathfrak{B}'$. }
\end{figure}

\section{Incoherent states}

Except for a class of particular frames, in the case of frame-dependent coherence, the set of all the incoherent states $\mathcal{I}$ is empty.
The frame-dependent version of coherence of a quantum state $\varrho$ is defined directy by using the matrix of  $\varrho$ in the 
chosen frame, without the need to use the set $\mathcal{I}$ of all the incoherent states.

If the Hilbert space $\mathcal{H}$ is the orthogonal sum $\mathcal{H}\!=\!\mathcal{H}_1\oplus \mathcal{H}_2$,
and if $\{ |\psi_1\rangle,  |\psi_2\rangle, ..., |\psi_n\rangle\}$ is an orthonormal basis in $\mathcal{H}_1$ and if 
$\{ |\varphi_1\rangle,  |\varphi_2\rangle, ..., |\varphi_m\rangle\}$ is a tight frame in $\mathcal{H}_2$, then
$\mathcal{F}\!=\!\{ |\psi_1\rangle,  |\psi_2\rangle, ..., |\psi_n\rangle, |\varphi_1\rangle,  |\varphi_2\rangle, ..., |\varphi_m\rangle\}$ is a tight frame in  $\mathcal{H}$,
\begin{equation}
\sum\limits_{j=1}^n|\psi_j\rangle\langle \psi_j|\!+\!\sum\limits_{k=1}^m |\varphi_k\rangle\langle \varphi_k|\!=\!\mathbb{I}_\mathcal{H}.
\end{equation}
In this case, any quantum state $\varrho$ of the form
\begin{equation}
\varrho\!=\!\sum\limits_{j=1}^n\lambda_j |\psi_j\rangle\langle \psi_j| 
\end{equation}
with $\lambda_j\geq0$ and $\sum_{j=1}^n\lambda_j\!=\!1$ is incoherent because $\langle \psi_j| \varrho|\psi_k\rangle \!=\!0$ for $j\!\neq\!k$, and $\langle \psi_j| \varrho|\varphi_k\rangle \!=\langle  \varphi_j| \varrho| \psi_k\rangle \!=\langle  \varphi_j| \varrho| \varphi_k\rangle \!=\!0$ for all $j,\, k$.

For example, with respect to the frame 
\begin{equation}
\mathfrak{F}\!=\!
{\small
  \left\{\left(\!\!
\begin{array}{r}
1\\
0\\
0
\end{array}\!\!
\right)\! , \left(\!\!
\begin{array}{r}
0\\
1\\
0
\end{array}\!\!
\right) \!,\frac{1}{2}\!\left(\!\!
\begin{array}{r}
0\\
0\\
1
\end{array}\!\!
\right)\!, \frac{\sqrt{3}}{2}\!\left(\!\!
\begin{array}{r}
0\\
0\\
1
\end{array}\!\!
\right)
\right\}
}
\end{equation}
of $\mathbb{C}^3$, the quantum states
\begin{equation}
\varrho (\alpha)\!=\!
\left( \!\!\begin{array}{ccc}
\alpha  & 0 & 0\\[1mm]
0 & 1\!-\!\alpha & 0\\[1mm]
0 & 0 & 0
\end{array} \!\!\right),\quad \mbox{where}\quad \alpha \!\in\![0,1],
\end{equation}
are incoherent states, that is $\mathcal{C}_{\mathfrak{F}}(\varrho(\alpha))\!=\!0$.

\section{Coherence with respect to a system of coherent states}

In the case of a quantum system described by the odd-dimensional  Hilbert space
\begin{equation}
\mathcal{H}\!=\!\{ \psi \!:\!\{-s,-s\!+\!1,...,s\!-\!1, s\}\longrightarrow \mathbb{C}\},\quad 
 \langle \varphi,\psi\rangle\!=\!\sum\limits_{k=-s}^{s}\overline{\varphi(k)}\, \psi(k)
\end{equation}
of dimension $d\!=\!2s\!+\!1$, the discrete version of the system of canonical coherent states is a remarkable tight frame.
Each function $ \psi \!\in\!\mathcal{H}$ is regarded as the restriction to $\{-s,-s\!+\!1,...,s\!-\!1, s\}$
of a periodic function $\psi \!:\!\mathbb{Z}\longrightarrow \mathbb{C}$ of period $d$.

For any $\kappa \!\in\!(0,\infty)$, the function
\begin{equation}
g_\kappa \!:\!\{-\!s,\!-s\!+\!1,...,s\!-\!1,s\}\!\rightarrow\!\mathbb{R},\quad 
g_\kappa(n)\!=\!\sum\limits_{m=-\infty}^\infty \mathrm{Exp}\left[-\frac{\kappa \pi}{d}(n+md)^2\right]
\end{equation}
represents \cite{Vourdas04,Ruzzi06,Cotfas11} a discrete version of the Gaussian function 
\begin{equation}
\mathbb{R}\!\rightarrow\!\mathbb{R}:q\mapsto  \mathrm{Exp}\left[-\frac{\kappa \pi}{h}q^2\right]
\end{equation}
and satisfies the relation \cite{Vourdas04,Ruzzi06,Cotfas11,Cotfas23}
\begin{equation}
Fg_\kappa\!=\!\mbox{${\small \frac{1}{\sqrt{\kappa}}}$}g_{\frac{1}{\kappa}},
\end{equation}
where $F\!:\!\mathcal{H}\!\rightarrow\!\mathcal{H}:\psi\!\mapsto\!F\psi$ is the Fourier transform
\begin{equation}
F\psi(k)\!=\!\mbox{${\frac{1}{\sqrt{d}}}$}\sum\limits_{j=-s}^s \mathrm{Exp}\left[-\frac{2\pi \mathrm{i}}{d}kj\right]\psi (j).
\end{equation}
Particularly, the normalized function  (see Fig.4)
\begin{equation}
|\mathbf{g}\rangle \!=\!\frac{1}{||g_1||}\, |g_1\rangle,
\end{equation}
satisfying $F|\mathbf{g}\rangle \!=\!|\mathbf{g}\rangle$, can be regarded as a {\em discrete version of the vacuum state}.  

\begin{figure}[t]
\includegraphics[scale=1.2]{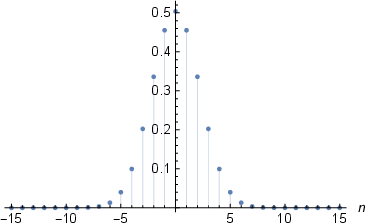}
\caption{Discrete version $\mathbf{g}$ of the vacuum state in the case $d\!=\!31$. }
\end{figure}

The $d^2$ {\em discrete coherent states}  \cite{Vourdas04,Ruzzi06,Cotfas11} 
\begin{equation}\label{discretecs}
|j,k\rangle \!=\!\frac{1}{\sqrt{d}}D(j,k)|\mathbf{g}\rangle,\qquad \mbox{where}\ \ j,k\!\in\!\{-\!s,\!-s\!+\!1,...,s\!-\!1,s\},
\end{equation}
defined by using the {\em displacement operators} 
$D(j ,k )\!:\!\mathcal{H}\!\rightarrow\!\mathcal{H},$
\begin{equation}
 D(j ,k )\psi (n)= \mathrm{Exp}\left[-\frac{\pi {\rm i}}{d}kj\right]\, \mathrm{Exp}\left[\frac{2\pi {\rm i}}{d}kn\right]\, \psi (n\!-\!j ),
\end{equation}
form a tight frame $\mathfrak{F}$  in $\mathcal{H},$
\begin{equation}
\sum\limits_{j,k=-s}^s|j,k\rangle \langle j,k|\!=\!\mathbb{I}.
\end{equation}
The coherence of a state $\varrho$ with respect to this tight frame is
\begin{equation}
\mathcal{C}_\mathfrak{F}(\varrho)\!=\!\frac{1}{d}\sum\limits_{
(j,k)\neq (n, m)
}|\langle j,k|\varrho|n , m\rangle|.
\end{equation}
Because $F|j,k\rangle\!=\!|\!-\!k,j\rangle$, we have 
\begin{equation}
\mathcal{C}_\mathfrak{F}(\varrho)\!=\!\frac{1}{d}\sum\limits_{
(j,k)\neq (n, m)
}|\langle -k,j|\varrho|\!-\!m,n\rangle|\!=\!\frac{1}{d}\sum\limits_{
(j,k)\neq (n, m)
}|\langle j,k|F^\dag\varrho F|n , m\rangle|\!=\!\mathcal{C}_\mathfrak{F}(F^\dag\varrho F)\, ,
\end{equation}
that is, any state $\varrho$ and its Fourier transform $F^\dag\varrho F$ have the same coherence.

In the three-dimensional case $\mathbb{C}^3\!\equiv\!\{ \psi \!:\!\{-1,\, 0,\, 1\}\longrightarrow \mathbb{C}\}$, the  vacuum state is 
\begin{equation}
\mathbf{g}\!:\!\{-1,\, 0,\, 1\}\longrightarrow \mathbb{C},\quad \mathbf{g}(0)\!=\!\mbox{$\frac{1}{\sqrt{2}}\sqrt{1\!+\!\frac{1}{\sqrt{3}}}$}, \quad \mathbf{g}(-1)\!=\!\mbox{$\frac{1}{2}\sqrt{1\!-\!\frac{1}{\sqrt{3}}}$}\!=\! \mathbf{g}(1).
\end{equation}
For example, the coherence of the  quantum state (\ref{rho3})   is $\mathcal{C}_\mathfrak{F}(\varrho)\!=\!1.259$.

\section{Frame-dependent versus POVM-based coherence}

A general measurement of a quantum system is described by a positive operator valued measure (POVM), that is,
by a set $\{ E_1,E_2,...,E_n\}$ of operators satisfying the conditions $E_j\geq0$ and $\sum_{j=1}^n E_j\!=\!\mathbb{I}$.
In this case, there exist some measurement operators  $\{ A_1,A_2,...A_n\}$ such that $E_j\!=\!A_j^\dag A_j$.
The probability to obtain the $j$-th outcome when measuring a quantum state $\varrho$ is $p_j\!=\!\mathrm{tr}(\varrho E_i)$,
and the $j$-th post-measurement state is $\varrho_j\!=\!\frac{1}{p_j}A_j\varrho A_j^\dag$.

If $\{ |\psi_1\rangle, |\psi_2\rangle,..., |\psi_n\rangle\}$ is a tight frame, then 
$\{ E_1\!=\!|\psi_1\rangle\langle\psi_1|, E_2\!=\!|\psi_2\rangle\langle\psi_2|, ..., E_n\!=\!|\psi_n\rangle\langle\psi_n|\}$  
is a POVM, and conversely, if all the elements of a POVM  $\{ E_1,E_2,...,E_n\}$ are rank-one operators,
 $E_j\!=\!|\psi_j\rangle\langle\psi_j|$, then  $\{ |\psi_1\rangle, |\psi_2\rangle,..., |\psi_n\rangle\}$ is a tight frame.
By using this identification, the tight frames can be regarded as a particular case of POVMs.
So, a POVM-based version of coherence can be defined in the case of any tight frame.

According to the Naimark theorem, by embedding the Hilbert $\mathcal{H}$ of the quantum system  into a higher-dimensional Hilbert space 
$\mathcal{H}'$, every POVM $\{ E_1,E_2,...,E_n\}$ can be extended to a projective measurement
$\{ \Pi_1,\Pi_2,...,\Pi_n\}$ on $\mathcal{H}'$ satisfying the conditions 
$\sum_{j=1}^n \Pi_j\!=\!\mathbb{I}_{\mathcal{H}'}$, $\Pi_j^2\!=\!\Pi_j$ and $\Pi_j\Pi_k\!=\!0$ for $j\!\neq\!k$.
Naimark extension is not unique. The most general way is via a direct sum by requering \cite{Decker05,Bischof19}
\begin{equation}
\mathrm{tr}(\varrho\, E_j)\!=\!\mathrm{tr}((\varrho\oplus 0)\Pi_j),\qquad \mbox{for any }\ j\!\in\!\{1,2,...,n\},
\end{equation}
where $0$ represents the zero matrix of dimension $\mathrm{dim}\, \mathcal{H}'\!-\!\mathrm{dim}\, \mathcal{H}$.

For example, the tight frame (\ref{triangular}),namely
\begin{equation}
\mathcal{F}_3\!=\!\left\{ |\psi_0\rangle\!=\!\left(\begin{array}{r}
\sqrt{\frac{2}{3}}\\
0
\end{array}\right), \, 
 |\psi_1\rangle\!=\!\left(\begin{array}{r}
-\frac{1}{\sqrt{6}}\\[1mm]
\frac{1}{\sqrt{2}}
\end{array}\right), \, 
 |\psi_2\rangle\!=\!\left(\begin{array}{r}
-\frac{1}{\sqrt{6}}\\[1mm]
-\frac{1}{\sqrt{2}}
\end{array}\right)
\right\} 
\end{equation}
identified with the POVM $\{E_0\!=\! |\psi_0\rangle\langle \psi_0|, E_1\!=\! |\psi_1\rangle\langle \psi_1|, E_2\!=\! |\psi_2\rangle\langle \psi_2|\}$ on $\mathbb{C}^2$ admits the Naimark extension to the projective measurement
$\{\Pi_0\!=\! |\Psi_0\rangle\langle \Psi_0|, \Pi_1\!=\! |\Psi_1\rangle\langle \Psi_1|, \Pi_2\!=\! |\Psi_2\rangle\langle \Psi_2|\}$,
where
\begin{equation}
\mathcal{B}\!=\!\left\{ |\Psi_0\rangle\!=\!\left(\begin{array}{r}
\sqrt{\frac{2}{3}}\\
0\\
\frac{1}{\sqrt{3}}
\end{array}\right), \, 
 |\Psi_1\rangle\!=\!\left(\begin{array}{r}
-\frac{1}{\sqrt{6}}\\[2mm]
\frac{1}{\sqrt{2}}\\[2mm]
\frac{1}{\sqrt{3}}
\end{array}\right), \, 
 |\Psi_2\rangle\!=\!\left(\begin{array}{r}
-\frac{1}{\sqrt{6}}\\[2mm]
-\frac{1}{\sqrt{2}}\\[2mm]
\frac{1}{\sqrt{3}}
\end{array}\right)
\right\} 
\end{equation}
is an orthonormal basis in $\mathbb{C}^3$. One can directly check that
\begin{equation}
\begin{array}{c}
\langle  \psi_j|\varrho |\psi_k\rangle \!=\!\langle  \Psi_j|\varrho\!\oplus \!0|\Psi_k\rangle,\\
\mathrm{tr}(\varrho E_j)\!=\!\mathrm{tr}((\varrho\!\oplus \!0)\Pi_j),
\end{array} 
\end{equation}
for any $j,k\!\in\!\{0,1,2\}$, and any quantum states
\begin{equation}
\varrho\!=\!
{\small
\left( \!\!\begin{array}{rr}
\varrho_{00}&\varrho_{01}\\
\varrho_{10}&\varrho_{11}
\end{array} \!\!\right)}\qquad \mbox{and}\qquad \varrho\!\oplus \!0\!=\!
{\small
\left( \!\!\begin{array}{rrr}
\varrho_{00}&\varrho_{01}& 0\\
\varrho_{10}&\varrho_{11} & 0\\
0 & 0 & 0
\end{array} \!\!\right)}.
\end{equation}

In the general case \cite{Decker05,Bischof19}, for a tight frame 
 $\mathcal{F}\!=\!\{ |\psi_1\rangle, |\psi_2\rangle,..., |\psi_n\rangle\}$ in $\mathbb{C}^d$, identified with the POVM
$\{ E_1\!=\!|\psi_1\rangle\langle\psi_1|, E_2\!=\!|\psi_2\rangle\langle\psi_2|, ..., E_n\!=\!|\psi_n\rangle\langle\psi_n|\}$,
we obtain a Naimark extension on  $\mathbb{C}^n$ by choosing 
$|\varphi_1\rangle,\,  |\varphi_2\rangle,\,...,\, |\varphi_n\rangle$ in $\mathbb{C}^{n-d}$ such that 
$\mathcal{B}\!=\!\{|\Psi_1\rangle,\, |\Psi_2\rangle,\, ...\, ,\, |\Psi_n\rangle\}$, where
\begin{equation}
|\Psi_j\rangle\!=\!|\psi_j\rangle\!\oplus\!|\varphi_j\rangle\equiv\left(\!\!\begin{array}{c}
|\psi_j\rangle\\
|\varphi_j\rangle
\end{array}\!\!\!\right),
\end{equation} 
is an orthonormal basis in $\mathbb{C}^n$. In this case, 
$\{\Pi_1\!=\! |\Psi_1\rangle\langle \Psi_1|, \Pi_2\!=\! |\Psi_2\rangle\langle \Psi_2|, ...,\Pi_n\!=\! |\Psi_n\rangle\langle \Psi_n|\}$
is a projective measurement on $\mathbb{C}^n$, and we have 
\begin{equation}
\mathrm{tr}((\varrho\!\oplus \!0)\Pi_j)\!=\!\langle \Psi_j|\varrho\!\oplus\!0|\Psi_j\rangle\!=\! \langle \psi_j|\varrho|\psi_j\rangle \!=\!\mathrm{tr}(\varrho E_j)
\end{equation} 
and the more general relation
\begin{equation}
\langle \Psi_j|\varrho\!\oplus\!0|\Psi_k\rangle\!=\!\left(\langle \psi_j|\ \ \langle \varphi_j|\, \right)
\left(\!\!\begin{array}{cc}
\varrho & 0\\
0 & 0
\end{array}\!\!\right)\left(\!\!\begin{array}{c}
|\psi_k\rangle\\
|\varphi_k\rangle
\end{array}\!\!\!\right)
\!=\!\langle \psi_j|\varrho|\psi_k\rangle ,
\end{equation} 
for any $j,k\!\in\!\{1,2,...,n\}$, and any quantum state 
\begin{equation}
\varrho\!=\!
{\small
\left( \!\!\begin{array}{ccc}
\varrho_{11}& \cdots & \varrho_{1d}\\
\cdots & \cdots & \cdots\\[-1mm]
\varrho_{d1}& \cdots &\varrho_{dd}
\end{array} \!\!\right)}\quad \mbox{and}\quad \varrho\!\oplus\! 0\!=\!
{\small
\left( \!\!\begin{array}{cccccc}
\varrho_{11}& \cdots & \varrho_{1d} & 0 &\cdots & 0\\
\cdots & \cdots & \cdots& \cdots & \cdots & \cdots\\[-1mm]
\varrho_{d1}& \cdots &\varrho_{dd} & 0 &\cdots & 0\\[1mm]
0 & \cdots & 0 & 0 & \cdots & 0 \\
\cdots & \cdots & \cdots& \cdots & \cdots & \cdots\\[-1mm]
0 & \cdots & 0 & 0 & \cdots & 0 
\end{array} \!\!\right)}.
\end{equation}
Particularly, we have 
\begin{equation}
\mathcal{C}_\mathfrak{F}(\varrho)\!=\!\frac{d}{n}\sum\limits_{j\neq k}|\langle \psi_j|\varrho|\psi_k\rangle|\!=\!\frac{d}{n}\sum\limits_{j\neq k}|\langle \Psi_j|\varrho\!\oplus\!0|\Psi_k\rangle| ,
\end{equation} 
that is, the  frame-dependent  coherence of $\varrho$ coincides (up to the multiplicative constant $d/n$) with the  basis-dependent $\ell_1$- norm of coherence of $\varrho\!\oplus\! 0$, in a Naimark extension. If we identify $\mathbb{C}^d$ to a subspace of $\mathbb{C}^n$ via the map
$\mathbb{C}^d\!\rightarrow\!\mathbb{C}^n\!:\!|x\rangle\!\mapsto\!|x\rangle\!\oplus\!|0\rangle$, then the frame $\mathfrak{F}$ is the orthogonal projection of the orthonormal basis $\mathfrak{B}$.

\section{Frame-dependent coherence of the states of a composite quantum system}

\begin{figure}[h]
a)\includegraphics[scale=1.0]{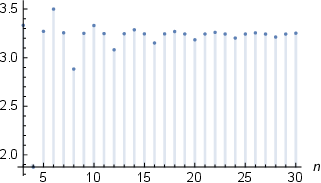}\qquad 
b)\includegraphics[scale=1.0]{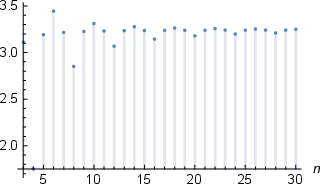}\\[5mm]
c)\includegraphics[scale=1.0]{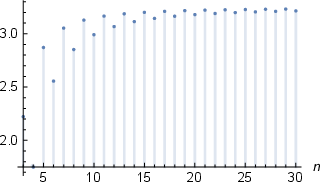}\qquad 
d)\includegraphics[scale=1.0]{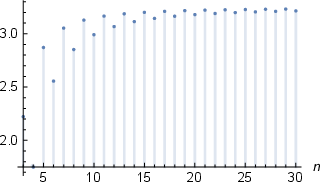}
\caption{Frame coherence of Bell states: a) $\Psi_1$, \  b) $\Psi_2$,
\ c) $\Psi_3$, \ d) $\Psi_4$ for $n\!\in\!\{ 3,4,5,...,30\}$. }
\end{figure}

The definition of the frame-dependent coherence can be extended in the usual way to composite quantum systems.
If $\mathcal{H}_{\!_A}$, $\mathcal{H}_{\!_B}$ are two Hibert spaces, $\mathrm{dim}\,\mathcal{H}_{\!_A}\!=\!d_A$,  $\mathrm{dim}\,\mathcal{H}_{\!_B}\!=\!d_B$, $\mathfrak{F}_{\!_A}\!=\!\{ |\varphi_1\rangle_{\!_A}, \, |\varphi_2\rangle_{\!_A},\, ...\, , \, |\varphi_{n_A}\rangle_{\!_A}\}$ is a frame of  $\mathcal{H}_{\!_A}$ and $\mathfrak{F}_{\!_B}\!=\!\{ |\varphi_1\rangle_{\!_B}, \, |\varphi_2\rangle_{\!_B},\, ...\, , \, |\varphi_{n_B}\rangle_{\!_B}\}$ is a frame of  $\mathcal{H}_{\!_B}$, then
\begin{equation}
\sum\limits_{j=1}^{n_A}\sum\limits_{k=1}^{n_B}|\varphi_j\varphi_k\rangle\langle\varphi_j\varphi_k|\!=\!\sum\limits_{j=1}^{n_A}|\varphi_j\rangle_{\!_A}{}_{_A}\!\langle\varphi_j|\!\otimes\!\sum\limits_{k=1}^{n_B}|\varphi_k\rangle_{\!_B}{}_{_B}\!\langle\varphi_k|\!=\!\mathbb{I},
\end{equation}
that is, $\{ \, |\varphi_j\varphi_k\rangle\!=\!|\varphi_j\rangle_{\!_A}\!\otimes\!|\varphi_k\rangle_{\!_B}\ | \ 1\!\leq\!j\!\leq\!n_A,\ \ 1\!\leq\!k\!\leq\!n_B\}$
is a tight frame in $\mathcal{H}_{\!_A}\!\otimes\!\mathcal{H}_{\!_B}$.
The coherence of a state $\varrho$ of the composite system with respect to this tight frame is
\begin{equation}
\mathcal{C}_{\mathfrak{F}_{\!_A},\mathfrak{F}_{\!_B}}(\varrho)\!=\!\frac{d_A\, d_B}{n_A\, n_B}\sum\limits_{(j,k)\neq (m,\ell)}|\langle\varphi_j\varphi_k|\varrho|\varphi_m\varphi_\ell\rangle|.
\end{equation}

In the case of a system of two qubits, we can choose $\mathfrak{F}_{\!_A}\!=\!\mathfrak{F}_{\!_B}\!=\!\mathfrak{F}_n$, defined by (\ref{polyframe}).\\
For example, as concern the Bell states:

- the coherence of $\Psi_1\!=\!\frac{1}{\sqrt{2}}(|00\rangle +|11\rangle) $ is (see Fig. 5a)
\begin{equation}
\mathcal{C}_{\mathfrak{F}_n,\mathfrak{F}_n}(|\Psi_1\rangle\langle\Psi_1|)\!=\!\frac{8}{n^4}\sum\limits_{(j,k)\neq(m,\ell)}\left|\cos\frac{2(j-k)\pi}{n}\,\cos\frac{2(m-\ell)\pi}{n}\right|;
\end{equation}

- the coherence of $\Psi_2\!=\!\frac{1}{\sqrt{2}}(|00\rangle -|11\rangle) $ is  (see Fig. 5b)
\begin{equation}
\mathcal{C}_{\mathfrak{F}_n,\mathfrak{F}_n}(|\Psi_2\rangle\langle\Psi_2|)\!=\!\frac{8}{n^4}\sum\limits_{(j,k)\neq(m,\ell)}\left|\cos\frac{2(j+k)\pi}{n}\,\cos\frac{2(m+\ell)\pi}{n}\right|;
\end{equation}

- the coherence of $\Psi_3\!=\!\frac{1}{\sqrt{2}}(|01\rangle +|10\rangle) $ is  (see Fig. 5c)
\begin{equation}
\mathcal{C}_{\mathfrak{F}_n,\mathfrak{F}_n}(|\Psi_3\rangle\langle\Psi_3|)\!=\!\frac{8}{n^4}\sum\limits_{(j,k)\neq(m,\ell)}\left|\sin\frac{2(j+k)\pi}{n}\,\sin\frac{2(m+\ell)\pi}{n}\right|;
\end{equation}

- the coherence of $\Psi_4\!=\!\frac{\mathrm{i}}{\sqrt{2}}(|01\rangle -|10\rangle) $ is  (see Fig. 5d)
\begin{equation}
\mathcal{C}_{\mathfrak{F}_n,\mathfrak{F}_n}(|\Psi_4\rangle\langle\Psi_4|)\!=\!\frac{8}{n^4}\sum\limits_{(j,k)\neq(m,\ell)}\left|\sin\frac{2(j-k)\pi}{n}\,\sin\frac{2(m-\ell)\pi}{n}\right|.
\end{equation}

The numerical data presented in Fig. 5 suggest that the coherence of $\Psi_1$ and $\Psi_2$, as well as of $\Psi_3$ and $\Psi_4$, is the same. Again, for $n$ large enough, the coherence is almost independent on $n$, and can be regarded as a frame-independent coherence.

\section{Concluding remarks}

In the particular case when the POVM corresponds to a tight frame, the POVM-based $\ell_1$-norm of coherence of a quantum state 
can be defined similar to the basis-dependent $\ell_1$-norm of coherence, without the use of a Naimark extension. The frame-dependent coherence defined in this way offers several advantages:
\begin{itemize}
\item[-]  a more accurate description than the basis-dependent description;

\item[-]  a definition simpler than the definition used in the case when the frame is regarded as a POVM;

\item[-]  the possibility to use a frame containing two or more orthogonal bases;

\item[-]  the possibility to investigate in more details how the basis-dependent coherence changes when we pass from one basis to another;

\item[-] it is more adequate when we have to compare the coherence of two quantum states;

\item[-]  in order to measure the coherence,  it offers the possibility to choose more experimentally accessible observables;

\item[-]  it is less sensitive to the frame change than the basis-dependent coherence under the basis change

\item[-]  in the case of qubit or multi-qubit, it is possible to define a frame-invariant coherence as the value (see Fig.1)
of the coherence with respect to the regular polygonal frame (\ref{polyframe}) obtained for a large value of $n$.
\end{itemize}

It is known \cite{Sperling18} that an alternative definition of the basis-dependent coherence can be obtained by choosing an orthonormal basis
in the real Hilbert space $\mathcal{A}(\mathcal{H})$ of Hermitian operators instead of choosing an orthonormal basis in
$\mathcal{H}$. Particularly, the coefficients of the representation of $\varrho$ in the orthonormal basis of displaced parity operators is the Wigner function of $\varrho$. The presented frame-dependent version of the coherence can be extended in the following way. By starting from any frame in $\mathcal{H}$ we can construct \cite{Cotfas23} a tight frame in $\mathcal{A}(\mathcal{H})$. The coefficients of the standard representation of $\varrho$ in such a frame can be regarded as a more
general version of the Wigner function  \cite{Cotfas23}, and used in the investigation of the coherence of $\varrho$.

\end{document}